\documentclass{elsart}

\usepackage{epsfig}

\begin{document}

\begin{frontmatter}



\title{Dynamics of the Landau-Pomeranchuk-Migdal Effect in
	Au+Au Collisions at $\sqrt{s}=200$~AGeV}

\author[Duke]{Thorsten Renk,}
\author[Duke,RBRC]{Steffen A. Bass,}
\author[VECC]{and Dinesh K. Srivastava}

\address[Duke]{Department of Physics, Duke University,
             Durham, NC 27708-0305}
\address[RBRC]{RIKEN BNL Research Center, Brookhaven National Laboratory,
        Upton, NY 11973, USA}
\address[VECC]{Variable Energy Cyclotron Centre,
             1/AF Bidhan Nagar, Kolkata 700 064, India}

\begin{abstract}
We study the role played by the Landau-Pomeranchuk-Midgal (LPM) effect
in relativistic collisions of hadrons and heavy nuclei, within a parton cascade
model.
 We find that the LPM effect strongly affects the
gluon multiplication due to radiation and considerably 
alters the space-time evolution
of the dynamics of the collision. It ensures a  multiplicity 
distribution of hadrons in aggreement with the experimental 
proton-proton data. We study the
production of single photons in relativistic heavy ion collisions 
and find that the inclusion of LPM suppression leads to a reduction in
the single 
photon yield at small and intermediate transverse momenta.
The parton cascade calculation of the single photon yield including
the LPM effect is shown to be  
in good agreement with the recent PHENIX data taken at the Relativistic 
Heavy-Ion Collider.
\end{abstract}

\end{frontmatter}

% main text

\label{}

\section{Introduction}

Collisions of heavy nuclei at relativistic energies are expected to
lead to formation of a deconfined phase of strongly interacting
nuclear matter, often referred to as a Quark-Gluon-Plasma (QGP).
Recent data from the Relativistic Heavy-Ion Collider (RHIC) at
Brookhaven Lab have provided strong indications for the existence of a
transient QGP -- among the most exciting findings are strong
(hydrodynamic) collective flow
\cite{HuKoHe,teaney,PHENIXv2,Esumi,STAR-v2,STAR-v2b},
the suppression of high-$p_T$ particles
\cite{Wa1,GyViWa,PHENIX-su,Adams:2003kv}
and evidence for parton recombination as hadronization mechanism
at intermediate transverse momenta
\cite{FMNB:03prl,FMNB:03prc,GreKoLe:03prl,Vo,Nonaka:2003hx}.

A variety of theoretical models has been formulated to
describe the observed phenomena, e.g., fluid dynamical models,
perturbative QCD scattering models, as well as models based on
parton saturation and statistical approaches.
Although these models, which all contain adjustable parameters,
have been fairly successful within their respective domains of
anticipated applicability, they all have certain limitations. For
example, fluid dynamics cannot address the transport phenomena
occurring prior to local equilibration of the produced matter,
and it must fail above a certain, though unknown, value of $p_T$.
Perturbative parton scattering models fail to describe the
physics of equilibration and the evolution of collective flow.

On quite general grounds, one expects that there is an intermediate regime
between ultra-hard processes for which the dynamics resembles a 
superposition of essentially independent nucleon collisions and soft 
processes which can be described by fluid dynamics  
In this regime, semi-hard rescattering of partons produced in primary hard 
collisions is important, but still perturbatively calculable.

% Initial state parton saturation models do not include the final
% state interaction among partons, which leads to equilibration.
% Seveveral of these issues are addressed 

This intermediate regime is addressed by the parton cascade model (PCM),
albeit in a semi-classical manner. 
The parton
cascade model \cite{GM} was proposed to provide a detailed
description of the temporal evolution of nuclear collisions at high
energy, from the onset of hard interactions among the partons of the
colliding nuclei up to the moment of hadronization. The PCM is based
on a relativistic Boltzmann equation for the time evolution of the
parton density in phase space due to perturbative QCD interactions
including scattering and radiation in the leading logarithmic
approximation.

In the present work we report on consequences of
Landau-Pomeranchuk-Midgal (LPM)
effect on the evolution of the collision within the framework of 
the PCM approach.

The LPM effect was first 
derived for QED \cite{LPM} and describes the destructive
interference between bremsstrahlung amplitudes in the case
of multiple scattering of the radiating particle. It in 
effect interpolates between the Bethe-Heitler and factorization
limits for the radiation spectrum of a charged particle 
undergoing multiple scatterings in a medium. The Bethe-Heitler
limit is obtained when the separation between individual
scattering centers becomes sufficiently large so that 
the radiation off these centers can be calculated 
as an incoherent sum  of radiation spectra resulting 
from the individual small-angle scatterings.
In the factorization limit the individual scattering centers
sit too close together to be resolved by the emitted photon.
The observed radiation then factorizes into a product of a 
single scattering radiation spectrum for the sum of the
momentum transfers obtained in all successive 
small angle scatterings of the emitter and its elastic cross 
section for this momentum transfer accumulated over these
small-angle scatterings. In the regime between those two
limiting cases, the LPM effect describes the suppression
of radiation relative to the Bethe-Heitler limit in regimes
where the radiation formation time is long compared to the 
mean free path of the emitter and thus destructive interference
between the bremsstrahlung amplitudes becomes important.
The QED LPM effect has recently been verified by experiments
at SLAC \cite{slac146}. 

Calculations have shown that the QCD analogue of
the LPM effect (it differs from QED due to the non-abelian
nature of QCD) also plays an important role, in 
particular for estimating the energy loss $dE/dx$ of an
energetic parton traversing a dense QCD medium 
\cite{GW,BDMPS,zakharov,urs}.
However, all these calculations have focussed on limiting
cases of infinitely many or very few ($N<3$) rescatterings
of the parton and have either been performed for a static
medium or have utilized only very schematic scenarios
(e.g. boost-invariant longitudinal expansion)
for the evolution of the partonic medium created in a 
heavy-ion collision. 

The LPM effect is expected to limit the growth of parton multiplicity
in the dense spacetime-regions of the scattering system. For a perturbative
framework like the PCM, this implies that the sensitivity to a soft cutoff
scale parameter for particle production is greatly reduced.

We note that other hard scattering models like PYTHIA \cite{Pythia}, the DPM \cite{DPM} and HIJING \cite{HIJING} contain only momentum 
space physics, they do not include any attempt to address spacetime dynamics. It is therefore 
difficult to investigate questions dealing with spacetime dynamics (like the LPM effect) 
within this class of models. All hard scattering models contain parameters associated with e.g. the 
separation of hard and soft scales - these parameters are adjusted such that the model 
describes the p-p data well (in the above cases without rescattering) - but the fact that 
this is possible for p-p collisions should not be taken as an indication that there 
is no rescattering or that it would not be important. 
In \cite{wang,klaus} is has been found that high multiplicity events correspond to underlying events 
with a high number of hard collisions - it is not unreasonable to suspect that such a high 
number in the small volume given by the p-p overlap will lead to rescattering (and LPM suppression).
In fact, it is the simultaneous description 
of hard p-p (where rescattering may be observed but is not dominant) and hard 
A-A physics (where rescattering is expected to be an integral part of the dynamics) with 
the same underlying set of parameters which is the decisive test for the importance of
rescattering and LPM suppression, and we aim to investigate this question in the following.

We describe a schematic
implementation of the LPM suppression of gluon and photon
radiation into a microscopic transport model, allowing us to 
study its effect on the non-equilibrium space-time-evolution
and reaction dynamics of a heavy-ion collision at RHIC. We show that the same
implementation of the LPM suppression is able to describe two sets of data which 
represent very different conditions in the collision system, i.e. the multiplicity distribution of produced
secondary particles in p-p collisions for different $\sqrt{s}$ and photon production 
in 200 AGeV Au-Au collisions. 
We select this particular choice of observables because there is no multiplicity distribution data 
available for heavy-ion collisions since the measured multiplicity is used to 
determine collision centrality. Direct photon emission is then the 
cleanest observable measuring the amount of hard collisions taking place in the pQCD rescattering phase.

We argue that this simultaneous agreement demonstrates that we have indeed been able to
introduce the physics of LPM supression into the PCM correctly and that we have 
a reliable description of the physics in the regime between soft fluid dynamics and
the hard pQCD regime where processes scale with the number of binary collisions.

\section{The Parton Cascade Model}

The fundamental assumption underlying the PCM is that the state
of the dense partonic system can be characterized by a set of
one-body distribution functions $F_i(x^\mu,p^\alpha)$, where $i$
denotes the flavor index ($i = g,u,\bar{u},d,\bar{d},\ldots$)
and $x^\mu, p^\alpha$ are coordinates in the eight-dimensional
phase space. The partons are assumed to be on their mass shell,
except before the first scattering. In our numerical implementation,
the GRV-HO parametrization \cite{grv} is used, and the parton
distribution functions are sampled at an initialization scale
$Q_0^2$ ($\approx (p_T^{\rm{min}})^2 $; see later) to create a discrete
set of particles. Partons generally propagate on-shell and along
straight-line trajectories between interactions. Before their first
collision, partons may have a space-like four-momentum, especially
if they are assigned an ``intrinsic'' transverse momentum.

The time-evolution of the parton distribution is governed by a
relativistic Boltzmann equation:
\begin{equation}
p^\mu \frac{\partial}{\partial x^\mu} F_i(x,\vec p) = {C}_i[F]
\label{eq03}
\end{equation}
where the collision term ${C}_i$ is a nonlinear functional
of the phase-space distribution function. The calculations discussed
below include all lowest-order QCD scattering processes between
massless quarks and gluons \cite{Cutler.78}, as well as all
($2 \to 2$) processes involving the emission of photons
($qg \to q\gamma$, ${\bar q}g \to {\bar q}\gamma$,
$q{\bar q} \to g\gamma$, $q{\bar q} \to \gamma\gamma$).
A low momentum-transfer cut-off $p_T^{\rm{min}}$ is needed to
regularize the infrared divergence of the perturbative parton-parton
cross sections. A more detailed description of our implementation is
in preparation \cite{bms_big1}.

We account for the final state radiation~\cite{Ben87} following a
hard interaction in the parton shower approach. In the leading
logarithmic approximation, this picture corresponds to a sequence
of nearly collinear $1 \to 2$ branchings: $a \to bc$. Here $a$ is
called the mother parton, and $b$ and $c$ are called daughters.
Each daughter can  branch again, generating a structure with a
tree-like topology. We include all such branchings allowed by
the strong and electromagnetic interactions.
The probability for a parton to branch is given in terms of the
variables $Q^2$ and $z$.  $Q^2$ is the momentum scale of the
branching and $z$ describes the distribution of the energy of the
mother parton $a$ among the daughters $b$ and $c$, such that $b$
takes the fraction $z$ and $c$ the remaining fraction $1-z$. The
differential probability to branch is:
\begin{equation}
{dP}_a=\sum_{b,c} \frac{\alpha_{abc}}{2\pi}P_{a \to bc}
           \frac{dQ^2}{Q^2} dz
\end{equation}
where the sum runs over all allowed branchings. The $\alpha_{abc}$ is
$\alpha_{em}$ for branchings involving emission of a photon and
$\alpha_s$ for the QCD branchings. The splitting kernels
$P_{a \rightarrow bc}$ are given in \cite{Alt77}. The collinear
singularities in the showers are regulated by terminating the
branchings when the virtuality of the (time-like) partons drops
to $\mu_0$, which we take as 1 GeV. We note that there is no
great sensitivity to the detailed choice of $\mu_0$ as soon as 
the LPM suppression is included (a reduction to $\mu_0$ = 0.5 GeV leads to less than
30\% change in parton production) since the LPM effect limits the density
of produced partons.  In principle, one could take
a smaller value for the cut-off $\mu_0$ for a quark fragmenting
into a photon~\cite{TS}, but we have not done so as
we are only interested in high energy photons here. The soft-gluon
interference is included as in \cite{Ben87}, namely by selecting
the angular ordering of the emitted gluons.
An essential difference between emission of a photon and a parton
in these processes is that the parton encounters further interactions
and contributes to the build-up of the cascade, while the photon
almost always (in our approximation, always) leaves the system
along with the information about the interaction.

Since the microscopic degrees of freedom of the PCM, quarks and
gluons, are treated as quasi-particles, a full quantum 
implemention of the LPM is beyond the scope of the model.
In order to take the main characteristics of the LPM effect into
account, we introduce a formation time for the radiated particle,
\begin{equation}
\tau\,=\,  \frac{\omega}{k^2_\perp}\quad,
\end{equation}
with $\omega$ the energy of the radiated particle and $k_\perp$
its transverse momentum with respect to the emitter. During the formation time, the emitted
particles (which we refer to as {\em shower}) do not interact 
(and are thus assigned zero cross section).
The shower emitter, however, may  rescatter and if this
occurs within
duration of the formation time of the emitted particles, the
shower is considered to by suppressed by the LPM effect and
is removed from the further evolution of the collision system. 
Preliminary results of the effects of the LPM suppression on single
photon production in the PCM have earlier been reported in~\cite{bms_phot}.

It should be noted that a recent novel implementation of the PCM model
\cite{greiner_xu04}, based on a stochastic implementation of the collision
term (and thus allowing for detailed-balance conserving three-particle 
collisions), requires a very different modeling of the LPM effect.
In that approach the LPM effect is introduced via
a lower momentum cut-off for the gluon emission rate, leading to a nearly
isotropic angular distribution for inelastic scatterings and thus
shorter thermalization times. The effects
of this particular implementation on photon radiation  and multiplicity
scaling at RHIC remain to be investigated.

\section{Scaling of multiplicity distribtions scaling in p-p collisions}

In \cite{KNO} it was  suggested that asymptotically the distribution of
 the multiplicity of produced particles in p-p collisions
$\langle n\rangle \sigma_n(s)$ is only a function of $n/\langle n \rangle$, 
where $\sigma_n(s)$ is the multiplicity distribution for given 
center of mass energy $\sqrt{s}$ and $\langle n \rangle$ is the mean
 multiplicity for this $s$. Thus, the probability distribution
$P(n/\langle n \rangle)$ for producing a given fraction of the mean
 multiplicity would asymptotically be a universal function
$\Psi(n/\langle n \rangle)$ independent of $\sqrt{s}$.

In view of this expectation, a large body of data has been accumulated on the
multiplicity distribution of hadrons in p-p collisions at several energies,
and a description of these have remained an important check on the models
of hadronic interactions.  Deviations from this (KNO) scaling have also been 
studied extentively and are most clearly seen in events having
 high multiplicity 
at higher center of mass energies (see e.g. \cite{UA5}).
 The high multiplicity events
in p-p collisions necessarily involve increased multiple scatterings and gluon
multiplications in a small spacetime volume, 
when studied within models involving scattering of partons
(see e.g. Ref.~\cite{wang} and \cite{klaus}). Thus they provide the most easily tractable arena to study the consequences of the LPM effect.

% Experimentally, this scaling law is fulfilled rather well for 
%$\sqrt{s} > 40$ GeV.
%We demonstrate this in Fig.~\ref{F-KNO} using ISR data from \cite{ISR}.

The PCM should reproduce these multiplicity distributions in order to be reliable.
However,
there is one important caveat when comparing with data: The PCM does not 
include hadronization, thus numbers of
produced partons in the PCM have to be compared with measured hadron numbers.
 In the following, we make the
assumption that the number of partons produced in a collision scales with the number of measured
hadrons, i.e. $N_{part} \propto N_{had}$. This assumption has often been made in PCM studies 
(see Ref.~\cite{novack}).

\begin{figure}[htb]
\begin{center}
\epsfig{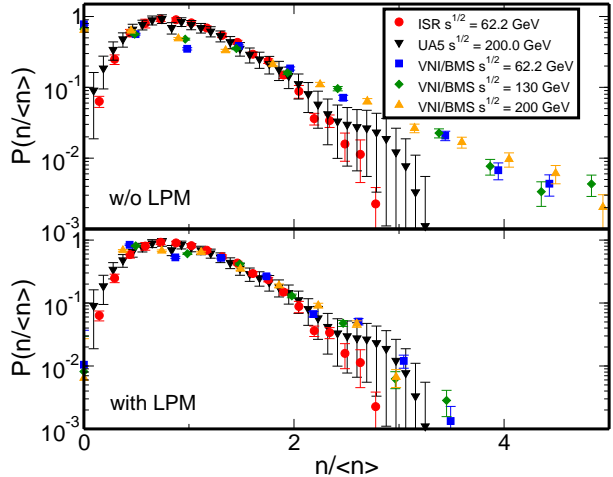}                                                                            
\caption{\label{F-KNO}Upper panel: Experimental data from \cite{ISR} and \cite{UA5} and PCM results without LPM reduction
for different $\sqrt{s}$ normalized as probability $P$ to find multiplicity $n$, plotted as a function
of the KNO scaling variable $n/\langle n \rangle$. Lower panel: As above, but PCM results include the
LPM suppression and non-single diffractive events only.}
\end{center}              
\end{figure}              

Under this assumption, we observe that the PCM without inclusion of
 the LPM 
suppression leads to a scaling of the multiplicity distribution in
$n/ \langle n \rangle$ (Fig.~\ref{F-KNO}  
upper panel). However the 'universal function' $\Psi^{PCM}$  is 
quite different from the 
measured $\Psi^{data}$. In particular, there is a large
probability to produce a high multiplicity. We note that in the region 
under investigation, i.e. $\sqrt{s} < 200$ GeV,
scaling violations are small and KNO scaling is fulfilled within 
experimental errors, i.e. the data show a 'universal function'.

%To some degree, this is even enhanced by the fact that the PCM allows events
%with zero particle production (which would not show up experimentally), this contribution leads to a lower $\langle n \rangle$ and
%thus widens $\Psi$ even more.
%
% As I said, the events of pcm with zero particle production correspond
% diffractive events, while these data are for NSD (non single 
%diffractive events). We need to plot them with the removal of i
%the diffractive events,.
%

Adjusting for the fact that we compare with non-single-diffractive events, we remove the events with zero particle production from 
the PCM event sample. Taking the effects of LPM suppression into
account, the resulting scaling function $\Psi^{PCM}$ is much closer to $\Psi^{data}$ and it is well conceivable that
hadronization can account for the remaining differences. This result gives some confidence that the implementation of the LPM
effect is done in a reasonable fashion. We have also confirmed that the PCM (including the LPM effect) 
is in agreement with the experimental multiplicity distribution for $\sqrt{s} = 900$ GeV 
\cite{UA5} with the same level of accuracy as in the comparison shown above.

\section{LPM dynamics in Au-Au collisions}

We now apply the same implementation of the LPM suppression  to 
Au-Au collisions where a correct treatment of the suppression is even more 
important due to the higher parton
density of the system. As a reference, we investigate Au-Au collisions at
 200 GeV/nucleon as realized 
at RHIC.

\begin{figure}[htb]
\begin{center}
\epsfig{file=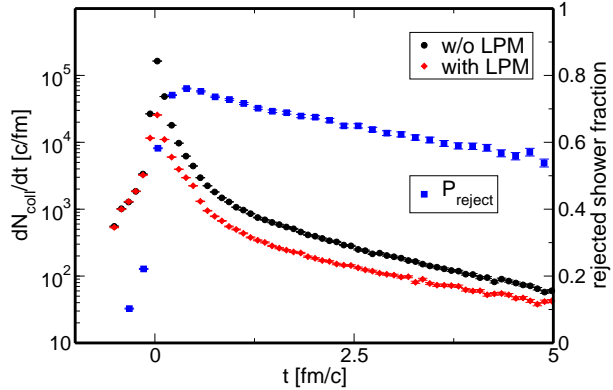, width=8cm}                                                                            
\caption{\label{F-collrate} Left axis: The collision rate in the PCM as 
a function of time for a scenario with (red) and without (black) LPM suppression. Right axis: The fraction of rejected showers as a function 
of time (blue).}
\end{center}              
\end{figure}      

In Fig.~\ref{F-collrate} we compare the collision rate as a function of time for the standard scenario (without LPM suppression)
and the one including the LPM effect. In both cases, the collision rate peaks strongly around maximum overlap of the
two nuclei at $t=0$ and then decreases rapidly as subsequent expansion dilutes the system. The LPM effect strongly
limits particle production in this high density peak, leading to a collision rate which is almost an order of
magnitude lower. Since particle production in the PCM proceeds by branching processes which create
soft partons, the result implies that the parton spectra remain harder if the LPM suppression is taken into account.

In addition we show the fraction $P_{\rm reject}$ of rejected showers as a function of time. Since in the PCM implementation of
the LPM effect the decision about shower rejection is made after a formation time $\tau=\omega/k^2_\perp$, the
maximum of the shower rejection does not coincide with the peak in the collision rate but is delayed. The result
indicates that the fraction of rejected showers is large throughout the whole evolution phase in which a perturbative
interaction picture is expected to be relevant.

\begin{figure}[htb]
\begin{center}
\epsfig{file=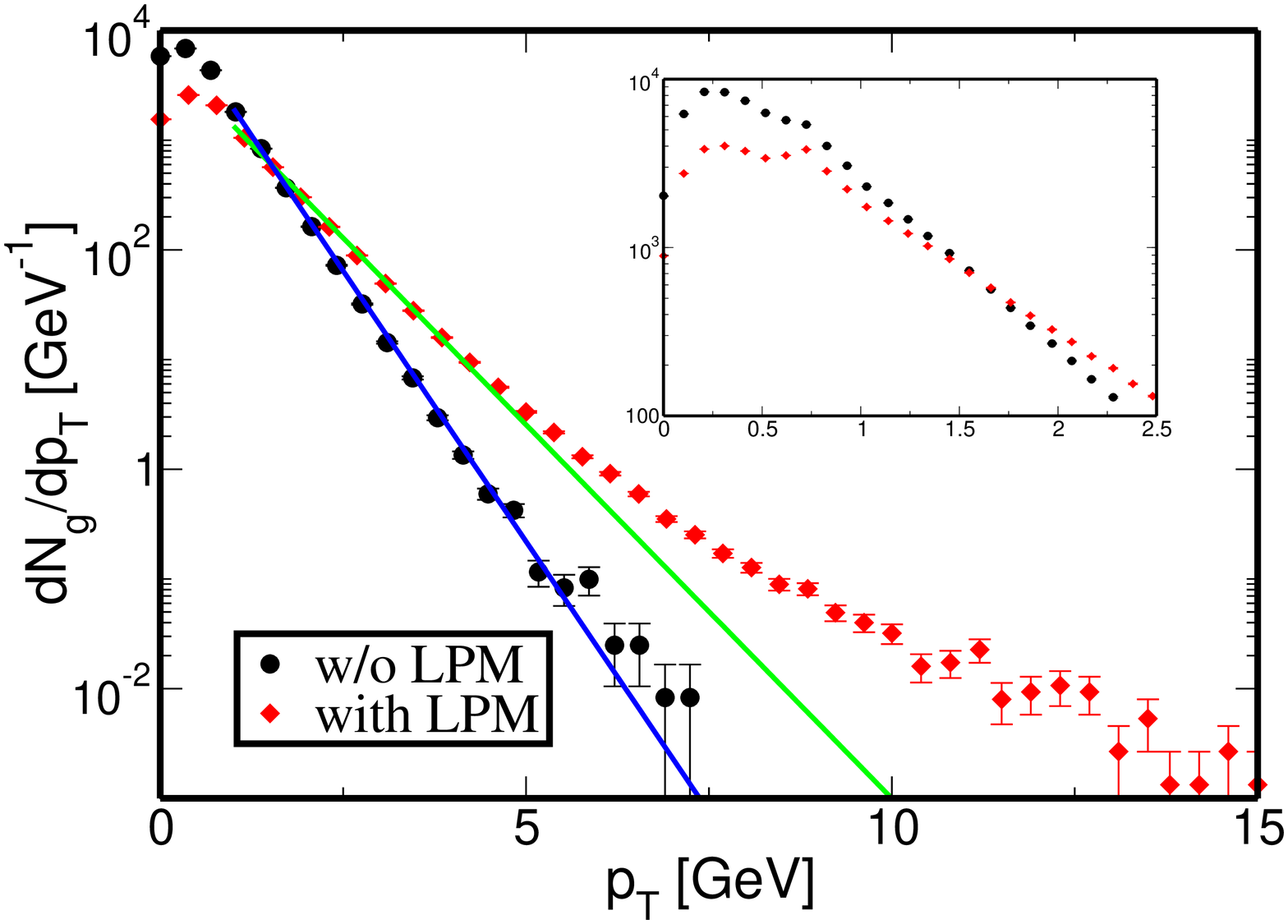, width=8cm}                                                                            
\caption{\label{F-g_spectra}Travsverse mass $p_T$ spectra of gluons for a scenario with (red) and without (black)
LPM suppression as obtained from the PCM. The insert shows the low $p_T$ region in greater detail. Shown are
also lines indicating exponential fits to the region 1 GeV $< p_T <$ 4 GeV of the spectra, corresponding
to apparent temperatures $T^\ast$ of 0.64 (0.44) GeV respectively. }
\end{center}              
\end{figure}      

This scenario is essentially confirmed by a direct comparison of gluon transverse momentum spectra in Fig.~\ref{F-g_spectra}.
The much higher collision rates in the standard scenario lead to an exponential spectrum with an apparent temperature $T^*$
(determined by a fit $\sim \exp[-p_T/T^*]$ in the range 1 GeV $< p_T <$ 4 GeV) of 0.44 GeV. In contrast, including the LPM
suppression leads to less spectral cooling (the apparent temperature is 0.64 GeV in the region where the spectrum is exponential)
and the power-law tail remains clearly visible for $p_T > 5$ GeV (note that $T^\ast$ is not a real temperature as
for example the longitudinal momentum spectrum looks very different from the transverse one and $dN/dE$ does not
follow the thermal distribution).
On the other hand, focusing on the low $p_T$ region 
reveals that including the LPM suppression leads to a factor $\sim 2.5$ reduction in the yield below $0.7$ GeV (the
$p_{T, \rm min}$ cutoff).

\section{Photon production in Au-Au collisions}

Ideally we would like to study a multiplicity scaling plot of Au-Au collisions to directly
compare to our p-p collision results. However, experimentally produced multiplicity is used to
determine the centrality class of the collision, hence a multiplicity distribution for central
Au-Au collisions is not available and we have to rely on a different probe sensitive to the 
number of hard collisions. Due to the smallness of the electromagnetic coupling, photons
produced in a heavy-ion collision can escape the interaction region essentially unaltered~\cite{Fei:76} 
and are therefore an excellent and reliable probe of the evolution of the partonic
cascade in nuclear collisions. In fact, one can argue that the
photons confirm the presence of parton cascading processes after
the initial primary parton-parton collisions. If we would attempt 
to describe p-p collisions without rescattering
processes and to carry the same description over to heavy-ion collisions
the resulting photon yield is reduced by about a factor of $\sim$ 3 -- 4 (cf. also the discussion in
\cite{bms_phot}).

Photons are produced in the PCM from Compton ($q g \rightarrow q \gamma$),
annihilation ($q \overline{q} \rightarrow g \gamma$), and
bremsstrahlung ($q^\star \rightarrow q \gamma$) processes.
These are analogous to
the processes governing the energy loss of energetic partons,
where gluons are emitted instead of photons.
As in \cite{bms_phot} we investigate the photon production during the
 hard initial 
stage of Au-Au collisions, focusing here on the effect of
the LPM suppression on single photon emission. 
We find a sizeable reduction of photon production
in particular at midrapidity (where experimental measurements have been mad) as compared to \cite{bms_phot} due to LPM suppression.

\begin{figure}[htb]
\begin{center}
\epsfig{file=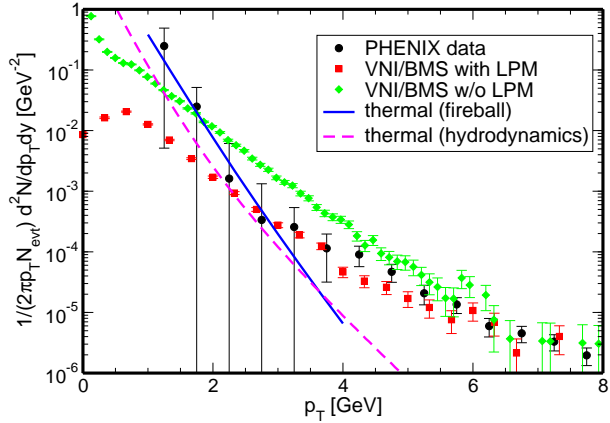, width=8cm}                                                                            
\caption{\label{F-gamma_spectra}
Photon transverse momentum spectra at midrapidity, as measured by the PHENIX collaboration \cite{PHENIX-gamma} and
as calculated in the PCM with (red) and without (green) inclusion of 
the LPM effect. Shown is also a calculation
of the thermal contribution to the spectrum using a fireball
model of expansion based on a fit to hadronic data (solid blue line),
 \cite{Photons1,Photons2}, and a hydrodynamic calculation 
(dashed magenta) \cite{Photons3}.
}
\end{center}              
\end{figure} 

Fig.~\ref{F-gamma_spectra} shows the  $p_T$ spectrum 
of single photons calculated with and without LPM suppression
as compared to the data obtained by the PHENIX
collaboration \cite{PHENIX-gamma} for the 0-10\% centralitiy class.
% The calculation including the LPM effect has been
%performed for central collisions and scaled with the ratio of
% the number of binary collisions
%to the measured 0-10\% centrality class.

We clearly observe that without the 
inclusion of the LPM suppression the
PCM overpredicts the data in the region 2 $< p_T < 3.5$ GeV whereas 
taking the LPM effect into account leads
to a fair description of the data. A thermal contribution to the 
photon yield, calculated using  a fireball model \cite{Photons1,Photons2}
which reproduces HBT correlations for pions, 
or using a boost-invariant hydrodynamics with transverse
 expansion~\cite{Photons3}, with state of the art rates for thermal 
emission of photons from quark and hadronic matter,  seems
to improve the agreement with data in the low $p_T$ region, 
although the errors are large here. 
Note that without the presence of the LPM effect,
frequent soft scattering and branching processes 
generate a lot of photons below 0.7 GeV
(= $p_T^{\rm{min}}$). These processes are strongly suppressed by the LPM effect,
explaining the difference between a rise of the low $p_T$ spectrum (without LPM)
and a drop (including LPM) in the PCM results.

\section{Conclusions}

We have studied the role played by the Landau-Pomeranchuk-Midgal (LPM) effect
in relativistic collisions of hadrons and heavy nuclei, 
within the framework of the Parton Cascade Model.
 We find that the LPM effect strongly affects the
gluon multiplication due to radiation and considerably 
alters the space-time evolution of the dynamics of the collision. 
In particular, it restricts the growth of multiplicity due to soft
parton production considerably and strongly reduces the sensitivity of
parton production to the detailed choice of a soft cutoff parameter $\mu_0$.
It ensures a  multiplicity 
distribution of hadrons in aggreement with the experimental data
in proton-proton reactions. Furthermore, we find that the
production of single photons in relativistic heavy ion collisions 
is strongly affected -- the inclusion of the
LPM suppression leads to a depletion of single photons at low and 
intermediate transverse momenta up to 4 GeV/c and brings the PCM
calculation into good agreement with the recent PHENIX data taken at 
the Relativistic Heavy-Ion Collider. The success in the reproduction of these
two different sets of data is far from trivial as they represent not only very
different observables (bulk production vs. rare process) but also vastly different
$\sqrt{s}$ and parton densities. This success gives some confidence that 
we have a useful description of the regime in which collisions are still perturbatively
calculable but multiple re-scattering is important.

\section*{Acknowledgments}
This work was supported in part by RIKEN, the Brookhaven National
Laboratory, DOE grants DE-FG02-05ER41367 and DE-AC02-98CH10886
and a Feodor Lynen Fellowship of the Alexander von Humboldt Foundation.

% \appendix

% \section{}
% \label{}


\begin{thebibliography}{00}

\bibitem{HuKoHe}
%RADIAL AND ELLIPTIC FLOW AT RHIC: FURTHER PREDICTIONS.
P.~Huovinen, P.~F.~Kolb, U.~W.~Heinz, P.~V.~Ruuskanen and  S.~A.~Voloshin,
Phys.\ Lett.\ B {\bf 503}, 58 (2001).
%-Print Archive: hep-ph/0101136

\bibitem{teaney}
D.~Teaney, J.~Lauret and E.~V.~Shuryak,
%``A hydrodynamic description of heavy ion collisions at the SPS and RHIC,''
[arXiv:nucl-th/0110037].
%%CITATION = NUCL-TH 0110037;%%

\bibitem{PHENIXv2}
%Elliptic Flow of Identified Hadrons in Au-Au Collisions at
%sqrt(s_NN) = 200 GeV
S.~S.~Adler {\it et al.} [PHENIX Collaboration],
hys.\ Rev.\ Lett.\  {\bf 91}, 182301 (2003).
%%CITATION = NUCL-EX 0305013;%%

\bibitem{Esumi}
%dentified charged particle azimuthal anisotropy in PHENIX at RHIC
S.~Esumi (for the PHENIX collaboration),
Nucl.\ Phys.\  A{\bf 715}, 599 (2003).
%nucl-ex/0210012 [abs, ps, pdf, other]

\bibitem{STAR-v2}
C.~Adler {\it et al.} [STAR Collaboration], Phys.\ Rev.\ Lett.\ {\bf 90},
032301 (2003); {\it ibid.} {\bf 89} 132301 (2002); {\it ibid.}
{\bf 87} 182301 (2001).

\bibitem{STAR-v2b}
%Particle dependence of azimuthal anisotropy and nuclear modification
%of particle production at moderate pT in Au+Au collisions at sqrt(snn) = 200 GeVJ.~Adams {\it etal.} (STAR Collaboration), nucl-ex/0306007.
J.~Adams {\it et al.}  [STAR Collaboration],
% ``Particle dependence of azimuthal anisotropy and nuclear modification of
%particle production at moderate p(T) in Au + Au collisions at  s(NN)**(1/2) =
%200-GeV,''
Phys.\ Rev.\ Lett.\  {\bf 92}, 052302 (2004).
%%CITATION = NUCL-EX 0306007;%%

\bibitem{Wa1}
%JET QUENCHING AND AZIMUTHAL ANISOTROPY OF LARGE P(T) SPECTRA
%IN NONCENTRAL HIGH-ENERGY HEAVY ION COLLISIONS.
X.~N.~Wang, Phys.\ Rev.\ C {\bf 63}, 054902 (2001).
%nucl-th/0009019

\bibitem{GyViWa}
%HIGH P(T) AZIMUTHAL ASYMMETRY IN NONCENTRAL A+A AT RHIC
M.~Gyulassy, I.~Vitev, X.~N.~Wang, Phys.\ Rev.\ Lett.\ {\bf 86}, 2537 (2001).
%nucl-th/0012092

\bibitem{PHENIX-su}
%SUPPRESSION OF HADRONS WITH LARGE TRANSVERSE MOMENTUM IN CENTRAL
%Au+Au COLLISIONS at S_NN = 130 GEV
K.~Adcox {\it et al.} [PHENIX Collaboration], Phys.\ Rev.\ Lett. {\bf 88},
022301 (2002).

\bibitem{Adams:2003kv}
J.~Adams {\it et al.}  [STAR Collaboration],
 %``Transverse momentum and collision energy dependence of high p(T) hadron
%suppression in Au + Au collisions at ultrarelativistic energies,''
Phys.\ Rev.\ Lett.\  {\bf 91}, 172302 (2003).
%%CITATION = NUCL-EX 0305015;%%

\bibitem{FMNB:03prl}
R.~J.~Fries, B.~M\"uller, C.~Nonaka and S.~A.~Bass,
Phys. Rev. Lett. {\bf 90}, 202303 (2003).
%%CITATION = NUCL-TH 0301087;%%

\bibitem{FMNB:03prc}
R.~J.~Fries, B.~M\"uller, C.~Nonaka and S.~A.~Bass,
%``Hadron production in heavy ion collisions: Fragmentation and  recombination from a dense parton phase,''
Phys.\ Rev.\ C {\bf 68}, 044902 (2003).
%%CITATION = NUCL-TH 0306027;%%

\bibitem{GreKoLe:03prl}
%PARTON COALESCENCE AND THE ANTIPROTON/PION ANOMALY AT RHIC
V.~Greco, C.~M.~Ko, and P.~L\'evai, Phys.\ Rev.\ Lett.\ {\bf 90}, 202303 (2003).                                                                                

\bibitem{Vo}
%Anisotropic flow
S.~A.~Voloshin, Nucl.\ Phys.\ A {\bf 715} (2003) 379.
%nucl-ex/0210014

\bibitem{Nonaka:2003hx}
C.~Nonaka, R.~J.~Fries and S.~A.~Bass,
 %``Elliptic flow of multi-strange particles: Fragmentation, recombination and
%hydrodynamics,''
Phys.\ Lett.\ B {\bf 583}, 73 (2004).
%%CITATION = NUCL-TH 0308051;%%

\bibitem{GM}
K.~Geiger and B.~M\"uller,
%``Dynamics of parton cascades in highly relativistic nuclear collisions,''
Nucl.\ Phys.\ B {\bf 369}, 600 (1992).
%%CITATION = NUPHA,B369,600;%%


\bibitem{LPM}
L.~D.~Landau and I.~Pomeranchuk,
  %``Electron Cascade Process At Very High-Energies,''
  Dokl.\ Akad.\ Nauk Ser.\ Fiz.\  {\bf 92}, 735 (1953);
  %%CITATION = DANKA,92,735;%%
L.~D.~Landau and I.~Pomeranchuk,
  %``Limits Of Applicability Of The Theory Of Bremsstrahlung Electrons And Pair
  %Production At High-Energies,''
  Dokl.\ Akad.\ Nauk Ser.\ Fiz.\  {\bf 92}, 535 (1953);
  %%CITATION = DANKA,92,535;%%
A.~B.~Migdal,
  %``Bremsstrahlung And Pair Production In Condensed Media At High-Energies,''
  Phys.\ Rev.\  {\bf 103}, 1811 (1956).
  %%CITATION = PHRVA,103,1811;%%


\bibitem{slac146}
P.~L.~Anthony {\it et al.},
  %``An Accurate measurement of the Landau-Pomeranchuk-Migdal effect,''
  Phys.\ Rev.\ Lett.\  {\bf 75}, 1949 (1995);
  %%CITATION = PRLTA,75,1949;%%
P.~L.~Anthony {\it et al.}  [SLAC-E-146 Collaboration],
  %``Bremsstrahlung suppression due to the LPM and dielectric effects in a
  %variety of materials,''
  Phys.\ Rev.\ D {\bf 56}, 1373 (1997)
  [arXiv:hep-ex/9703016].
  %%CITATION = HEP-EX 9703016;%%

\bibitem{GW}
M.~Gyulassy and X.~n.~Wang,
  %``Multiple collisions and induced gluon Bremsstrahlung in QCD,''
  Nucl.\ Phys.\ B {\bf 420}, 583 (1994);
  %%CITATION = NUCL-TH 9306003;%%
X.~N.~Wang, M.~Gyulassy and M.~Plumer,
  %``The LPM effect in QCD and radiative energy loss in a quark gluon plasma,''
  Phys.\ Rev.\ D {\bf 51}, 3436 (1995).
  %%CITATION = HEP-PH 9408344;%%

\bibitem{BDMPS}
  R.~Baier, Y.~L.~Dokshitzer, A.~H.~Mueller, S.~Peigne and D.~Schiff,
  %``The Landau-Pomeranchuk-Migdal effect in QED,''
  Nucl.\ Phys.\ B {\bf 478}, 577 (1996);
  %%CITATION = HEP-PH 9604327;%%
R.~Baier, Y.~L.~Dokshitzer, A.~H.~Mueller, S.~Peigne and D.~Schiff,
  %``Radiative energy loss of high energy quarks and gluons in a  finite-volume
  %quark-gluon plasma,''
  Nucl.\ Phys.\ B {\bf 483}, 291 (1997);
  %%CITATION = HEP-PH 9607355;%%
R.~Baier, Y.~L.~Dokshitzer, A.~H.~Mueller and D.~Schiff,
  %``Medium-induced radiative energy loss: Equivalence between the BDMPS and
  %Zakharov formalisms,''
  Nucl.\ Phys.\ B {\bf 531}, 403 (1998);
  %%CITATION = HEP-PH 9804212;%%
R.~Baier, Y.~L.~Dokshitzer, A.~H.~Mueller and D.~Schiff,
  %``Radiative energy loss of high energy partons traversing an expanding  {QCD}
  %plasma,''
  Phys.\ Rev.\ C {\bf 58}, 1706 (1998).
  %%CITATION = HEP-PH 9803473;%%


\bibitem{zakharov}
B.~G.~Zakharov,
  %``Fully quantum treatment of the Landau-Pomeranchuk-Migdal effect in QED  and
  %QCD,''
  JETP Lett.\  {\bf 63}, 952 (1996);
  %%CITATION = HEP-PH 9607440;%%
B.~G.~Zakharov,
  %``Radiative energy loss of high energy quarks in finite-size nuclear  matter
  %and quark-gluon plasma,''
  JETP Lett.\  {\bf 65}, 615 (1997)
  [arXiv:hep-ph/9704255].
  %%CITATION = HEP-PH 9704255;%%

\bibitem{urs}
  U.~A.~Wiedemann and M.~Gyulassy,
  %``Transverse momentum dependence of the Landau-Pomeranchuk-Migdal effect,''
  Nucl.\ Phys.\ B {\bf 560}, 345 (1999).
  %%CITATION = HEP-PH 9906257;%%

 \bibitem{Pythia}
  T.~Sjostrand, P.~Eden, C.~Friberg, L.~Lonnblad, G.~Miu, S.~Mrenna and E.~Norrbin,
  %``High-energy-physics event generation with PYTHIA 6.1,''
  Comput.\ Phys.\ Commun.\  {\bf 135} (2001) 238.

\bibitem{DPM}
  A.~Capella, U.~Sukhatme and J.~Tran Thanh Van,
  Z.\ Phys.\ C {\bf 3}, 329 (1979);
  A.~Capella and J.~Tran Thanh Van,
  Z.\ Phys.\ C {\bf 10}, 249 (1981).
A.~Capella, U.~Sukhatme, C.~I.~Tan and J.~Tran Thanh Van,
  Phys.\ Rept.\  {\bf 236}, 225 (1994).


\bibitem{HIJING}
  M.~Gyulassy and X.~N.~Wang,
  %``HIJING 1.0: A Monte Carlo program for parton and particle production in
  %high-energy hadronic and nuclear collisions,''
  Comput.\ Phys.\ Commun.\  {\bf 83} (1994) 307.
   
\bibitem{wang} X.-N. Wang and M. Gyulassy, Phys. Rev. D {\bf 45} (1992) 844.

\bibitem{klaus} K. Geiger, Phys. Rev. D {\bf 47} (1993) 133.                                                                            
\bibitem{grv}
M.~Gl\"uck, E.~Reya and A.~Vogt,
%``Dynamical parton distributions of the proton and small x physics,''
Z.\ Phys.\ C {\bf 67}, 433 (1995).
%%CITATION = ZEPYA,C67,433;%%

\bibitem{Cutler.78}
R.~Cutler and D.~W.~Sivers,
%``Quantum Chromodynamic Gluon Contributions To Large P(T) Reactions,''
Phys.\ Rev.\ D {\bf 17}, 196 (1978);
%%CITATION = PHRVA,D17,196;%%
B.~L.~Combridge, J.~Kripfganz and J.~Ranft,
%``Hadron Production At Large Transverse Momentum And QCD,''
Phys.\ Lett.\ B {\bf 70}, 234 (1977).
%%CITATION = PHLTA,B70,234;%%

\bibitem{bms_big1}
S. A. Bass, B. M\"uller, and D, K, Srivastava, Phys. Lett. B {\bf 551}
(2003) 277; Phy. Rev. Lett. {\bf 91} (2003) 052302; J. Phys. G {\bf 29} (2003)
L 58; Phys. Rev. C {\bf 66} (2002) 061902 (R); and manuscript under preparation.

\bibitem{Ben87}
M. Bengtsson and T. Sj\"{o}strand,
Phys. Lett, B {\bf 185}, 435 (1987),
Nucl. Phys. B {\bf 289}, 810 (1987).
                                                                                
\bibitem{Alt77}
G. Altarelli and G. Parisi,
Nucl. Phys. B {\bf 126}, 298 (1977).

\bibitem{TS}
T. Sj\"{o}strand,
presented at the Workshop on Photon Radiation
from Quarks, Annency, 1991, Report. No. CERN-TH-6369/92.

\bibitem{bms_phot}
S.~A.~Bass, B.~Muller and D.~K.~Srivastava,
  %``Light from cascading partons in relativistic heavy-ion collisions,''
  Phys.\ Rev.\ Lett.\  {\bf 90}, 082301 (2003).
  %%CITATION = NUCL-TH 0209030;%%

\bibitem{greiner_xu04}
Z.~Xu and C.~Greiner,
  %``Thermalization of gluons in ultrarelativistic heavy ion collisions by
  %including three-body interactions in a parton cascade,''
  hep-ph/0406278.
  %%CITATION = HEP-PH 0406278;%%

\bibitem{KNO}
Z.~Koba, H.~B.~Nielsen and P.~Olesen, 
Nucl.\ Phys.\ B {\bf 40} (1972) 317.

\bibitem{ISR}
A.~Breakstone {\it et al.}  [Ames-Bologna-CERN-Dortmund-Heidelberg-Warsaw
                  Collaboration],
Phys.\ Rev.\ D {\bf 30} (1984) 528.


\bibitem{UA5}
G.~J.~Alner {\it et al.}  [UA5 Collaboration],
Phys.\ Lett.\ B {\bf 137} (1984) 304; 
Phys.\ Lett.\ B {\bf 167} (1986) 476.




\bibitem{novack} V. Borchers, J. Meyer, S. Gieseke, G. Martens, and 
C. C. Noack, Phys.\ Rev. \ C {\bf 62} (2000) 064903.


\bibitem{Fei:76}
E.~L.~Feinberg,
Nuovo Cim.\ {\bf 34A}, 391 (1976).

\bibitem{PHENIX-gamma}
  S.~S.~Adler {\it et al.}  [PHENIX Collaboration], nucl-ex/0503003.

\bibitem{Photons1}
T.~Renk, hep-ph/0408218.

\bibitem{Photons2}
  T.~Renk,
  %``Photonic measurements of the longitudinal expansion dynamics in  heavy-ion
  %collisions,''
  Phys.\ Rev.\ C {\bf 71} (2005) 064905

\bibitem{Photons3}
D. K. Srivastava, Phys. Rev. C {\bf 71} (2005) 034905.

\end{thebibliography}
\end{document}